\input{aipcheck}

\documentclass[
    ,final            
  ]
  {aipproc}

\layoutstyle{6x9}
\newcommand{\dhd}{{\textstyle d}
\lower.03ex\hbox{\kern-0.40em$^{\scriptstyle-}$}\kern-0.08em{}}

\begin{document}

\begin{flushright}
JLAB-THY-08-914\\
\end{flushright}

\title{NLO Evolution of Color Dipoles in $\mathcal{N}=4$ SYM\footnote{Talk given at the International
    Workshop on Diffraction in High-Energy Physics (Diffraction 2008)
    in La Londe-les-Maures, France, based on \cite{nlobk, nlobkn4}.}}


\author{Giovanni A. Chirilli}{
  address={Physics Dept, Old Dominion Univ.,
Norfolk, VA 23529, and\\
Theory Group, JLab, 12000 Jefferson Ave, Newport News, VA 23606\\
E-mail: chirilli@jlab.org}
}

\begin{abstract}

The small-$x_B$ deep inelastic scattering in the saturation region is governed by 
the non-linear evolution of Wilson-line operators. 
In the leading logarithmic approximation it is given by the BK
equation for the evolution of color dipoles. I discuss recent calculation of the 
next-to-leading order evolution of color dipoles in QCD and ${\cal N}=4$ SYM.
\end{abstract}

\maketitle


\section{Small-$x_B$ evolution of color dipoles}
A general feature of high-energy scattering is that a fast particle 
moves along its straight-line classical trajectory and the only quantum 
effect is the eikonal phase factor acquired along this propagation path. 
In QCD, for the fast quark or gluon scattering off some target, 
this eikonal phase factor is a Wilson line - the infinite gauge link ordered 
along the straight line collinear to the particle's velocity $n^\mu$:
\begin{equation}
U^\eta(x_\perp)={\rm Pexp}\Big\{ig\int_{-\infty}^\infty\!\!  du ~n_\mu 
~A^\mu(un+x_\perp)\Big\}~.~~~~
\label{defU}
\end{equation}
Here $A_\mu$ is the gluon field of the target, $x_\perp$ is the transverse
position of the particle which remains unchanged throughout the collision, and the 
index $\eta$ labels the rapidity of the particle. Repeating the above argument for the target (moving fast in the spectator's frame) we see that 
particles with very different rapidities perceive each other as Wilson lines and
therefore these Wilson-line operators form
the convenient effective degrees of freedom in high-energy QCD (for a review, see Ref.~\cite{mobzor}).
Let us consider the deep inelastic scattering from a hadron at small 
$x_B=Q^2/(2p\cdot q)$.  The virtual photon decomposes into a pair of fast quarks 
 moving along straight lines separated by some transverse distance.
The propagation of this quark-antiquark pair reduces  to the 
``propagator of the color dipole''  $U(x_\perp)U^\dagger(y_\perp)$ - two Wilson lines ordered 
along the direction collinear to the quarks' velocity. 
The structure function of a hadron is proportional to a matrix element of this color dipole operator
\begin{equation}
\hat{\cal U}^\eta(x_\perp,y_\perp)=1-{1\over N_c}
{\rm Tr}\{\hat{U}^\eta(x_\perp)\hat{U}^{\dagger\eta}(y_\perp)\}~,
\label{fla1}
\end{equation}
switched between the target's states ($N_c=3$ for QCD).   (As usual, we denote operators by ``hat'').
The energy dependence of the structure function is translated then into the dependence 
of the color dipole on the slope of the Wilson lines determined by the rapidity $\eta$.
Therefore, the  small-$x_B$ behavior of the structure functions for 
deep inelastic scattering from a hadron is  governed by the 
rapidity evolution of color dipoles. 
At relatively high energies and for sufficiently small dipoles we can use the leading logarithmic approximation (LLA)
where  $ \alpha_s\ll 1,~ \alpha_s\ln x_B\sim 1$ and get the non-linear BK evolution equation for the color
dipoles~\cite{npb96,yura}:
\begin{eqnarray}
&&\hspace{-3mm}
{d\over d\eta}~\hat{\cal U}(x,y)~=
\label{bk}\\
&&\hspace{-3mm}
{\alpha_sN_c\over 2\pi^2}\!\int\!d^2z~ {(x-y)^2\over(x-z)^2(z-y)^2}
[\hat{\cal U}(x,z)+\hat{\cal U}(y,z)-\hat{\cal U}(x,y)-\hat{\cal U}(x,z)\hat{\cal U}(z,y)]~.
\nonumber
\end{eqnarray}

The first three terms correspond to the linear BFKL evolution~\cite{bfkl} and describe the parton emission while the last term is responsible for the parton annihilation. For sufficiently high $x_B$ the parton emission balances the parton annihilation so the partons reach the state of saturation~\cite{saturation} with
the characteristic transverse momentum $Q_s$ growing with energy $1/x_B$. (For a review, see~\cite{satreviews}.)
\section{NLO evolution of color dipoles in QCD and $\mathcal{N}=4$ SYM}
As usual, to get the region of application of the leading-order evolution equation one needs to find the next-to-leading order (NLO) corrections. Another reason is that unlike the DGLAP evolution, the argument of the coupling constant in Eq. (\ref{bk}) is left undetermined in 
the LLA, and usually it is set by hand to be $Q_s$. Careful analysis of this argument is very important  from both theoretical and experimental points of view. 
Let us present the final result for the  NLO evolution of the color dipole~\cite{nlobkn4}
\begin{eqnarray}
&&\hspace{-2mm}
{d\over d\eta}{\rm Tr}\{\hat{U}_x \hat{U}^{\dagger}_y\}~
 =~{\alpha_s\over 2\pi^2}
\!\int\!d^2z~
{(x-y)^2\over X^2 Y^2}
\label{nlobk}\\
&&\hspace{-2mm}
\times~\Big\{1+{\alpha_s\over 4\pi}\Big[b\ln(x-y)^2\mu^2
-b{X^2-Y^2\over (x-y)^2}\ln{X^2\over Y^2}+
({67\over 9}-{\pi^2\over 3})N_c-{10\over 9}n_f
\nonumber\\
&&\hspace{-2mm} 
-~
2N_c\ln{X^2\over(x-y)^2}\ln{Y^2\over(x-y)^2}\Big]\Big\}
~[{\rm Tr}\{\hat{U}_x \hat{U}^{\dagger}_z\}{\rm Tr}\{\hat{U}_z \hat{U}^{\dagger}_y\}
-N_c{\rm Tr}\{\hat{U}_x \hat{U}^{\dagger}_y\}]   
\nonumber\\
&&\hspace{-2mm} 
+~{\alpha_s^2\over 16\pi^4}
\int \!d^2 zd^2 z'
\Bigg[
\Big(-{4\over (z-z')^4}+\Big\{2{X^2{Y'}^2+{X'}^2Y^2-4(x-y)^2(z-z')^2\over  (z-z')^4[X^2{Y'}^2-{X'}^2Y^2]}\nonumber\\ 
&&\hspace{-2mm}
+~{(x-y)^4\over X^2{Y'}^2-{X'}^2Y^2}\Big[
{1\over X^2{Y'}^2}+{1\over Y^2{X'}^2}\Big]
\nonumber\\ 
&&\hspace{-2mm}
+~{(x-y)^2\over (z-z')^2}\Big[{1\over X^2{Y'}^2}-{1\over {X'}^2Y^2}\Big]\Big\}
\ln{X^2{Y'}^2\over {X'}^2Y^2}\Big)
\nonumber\\ 
&&\hspace{-2mm}
\times~[{\rm Tr}\{\hat{U}_x\hat{U}^\dagger_z\}{\rm Tr}\{\hat{U}_z\hat{U}^\dagger_{z'}\}{\rm Tr}\{\hat{U}_{z'}\hat{U}^\dagger_y\}
-{\rm Tr}\{\hat{U}_x\hat{U}^\dagger_z \hat{U}_{z'}U^\dagger_y\hat{U}_z\hat{U}^\dagger_{z'}\}-(z'\rightarrow z)]
\nonumber\\ 
&&\hspace{-2mm}
+~\Big\{{(x-y)^2\over (z-z')^2 }\Big[{1\over X^2{Y'}^2}+{1\over Y^2{X'}^2}\Big]
-{(x-y)^4\over  X^2{Y'}^2{X'}^2Y^2}\Big\}\ln{X^2{Y'}^2\over {X'}^2Y^2}
\nonumber\\ 
&&\hspace{-2mm}
~{\rm Tr}\{\hat{U}_x \hat{U}^\dagger_z\}{\rm Tr}\{\hat{U}_z\hat{U}^\dagger_{z'}\}{\rm Tr}\{\hat{U}_{z'}\hat{U}^\dagger_y\}
\nonumber\\
&&\hspace{-2mm}
+~4n_f
\Big\{{4\over(z-z')^4}
-2{{X'}^2Y^2+{Y'}^2X^2-(x-y)^2(z-z')^2\over (z-z')^4(X^2{Y'}^2-{X'}^2Y^2)}
\ln{X^2{Y'}^2\over {X'}^2Y^2}\Big\}
\nonumber\\ 
&&\hspace{-2mm}
\times~{\rm Tr}\{t^a\hat{U}_xt^b\hat{U}^{\dagger}_y\}
[{\rm Tr}\{t^a
\hat{U}_zt^b \hat{U}^\dagger_{z'}\}-(z'\rightarrow z)]\Bigg]~.
\nonumber
\end{eqnarray}
Here $\mu$ is the normalization point in the $\overline{MS}$ scheme and
$b={11\over 3}N_c-{2\over 3}n_f$ is the first coefficient of the $\beta$-function (the quark part of Eq. (\ref{nlobk}) proportional to $n_f$ was found earlier~\cite{prd75,kw1}). 
The NLO kernel is a sum of the running-coupling part (proportional to $b$), the non-conformal  double-log 
term $\sim \ln{(x-y)^2\over (x-z)^2} \ln{(x-y)^2\over (x-z)^2}$ and the three conformal terms which depend on the two four-point conformal ratios ${X^2{Y'}^2\over {X'}^2Y^2}$ 
and ${(x-y)^2(z-z')^2\over X^2 {Y'}^2}$. Note that the logarithm of the second conformal ratio 
$\ln{(x-y)^2(z-z')^2\over X^2 {Y'}^2}$ is absent.   
The analysis of the argument of the coupling constant was performed in Refs.~\cite{prd75,kw1}. 
 It is possible to compare linearized NLO BK equation (\ref{nlobk}) with NLO BFKL in the case of 
forward scattering. The result (\ref{nlobk}) is in agreement with NLO BFKL equation~\cite{nlobfkl}  up to
a term proportional $\alpha_s^2\zeta(3)$ times the original dipole. 
We think that the difference could be due to different definitions of the cutoff in the longitudinal momenta.

In ${\cal N}=4$ SYM theory we have two additional types of diagrams: with scalar loops and with gluino loops.  
Let us present here the final result for the  NLO evolution of the color dipole in ${\cal N}=4$ SYM:
\begin{eqnarray}
&&\hspace{-2mm}
{d\over d\eta}{\rm Tr}\{\hat{U}_x \hat{U}^{\dagger}_y\}~
\label{nlobksym}\\
&&\hspace{-2mm} 
=~{\alpha_s\over 2\pi^2}
\!\int\!d^2z~
{(x-y)^2\over X^2 Y^2}\Big\{1+
{\alpha_sN_c\over 4\pi}\Big[
{1-\pi^2\over 3}
-~
2\ln{X^2\over(x-y)^2}\ln{Y^2\over(x-y)^2}\Big]\Big\}
\nonumber\\
&&\hspace{32mm} 
\times~[{\rm Tr}\{\hat{U}_x \hat{U}^{\dagger}_z\}{\rm Tr}\{\hat{U}_z \hat{U}^{\dagger}_y\}
-N_c{\rm Tr}\{\hat{U}_x \hat{U}^{\dagger}_y\}]   
\nonumber\\
&&\hspace{-2mm} 
+~{\alpha_s^2\over 16\pi^4}
\int \!d^2 zd^2 z'
\Big\{{(x-y)^4\over[ X^2{Y'}^2-{X'}^2Y^2] X^2{Y'}^2}
+{(x-y)^2\over (z-z')^2}{1\over X^2{Y'}^2}\Big\}
\ln{X^2{Y'}^2\over {X'}^2Y^2}
\nonumber\\ 
&&\hspace{-2mm}
\times~[{\rm Tr}\{\hat{U}_x\hat{U}^\dagger_z\}{\rm Tr}\{\hat{U}_z\hat{U}^\dagger_{z'}\}{\rm Tr}\{\hat{U}_{z'}\hat{U}^\dagger_y\}
-{\rm Tr}\{\hat{U}_x\hat{U}^\dagger_z \hat{U}_{z'}U^\dagger_y\hat{U}_z\hat{U}^\dagger_{z'}\}
\nonumber\\ 
&&\hspace{52mm}
-~{\rm Tr}\{\hat{U}_x\hat{U}^\dagger_{z'} \hat{U}_zU^\dagger_y\hat{U}_{z'}\hat{U}^\dagger_z\}-(z'\rightarrow z)]~.
\nonumber
\end{eqnarray}
The scalar and gluino contributions cancel two of the four terms in the QCD kernel so
the ${\cal N}=4$ NLO kernel is a sum of the  non-conformal  double-log 
term $\sim \ln{(x-y)^2\over (x-z)^2} \ln{(x-y)^2\over (x-z)^2}$ multiplied by the 
LO 1$\rightarrow$2 dipoles color structure and the  conformal term describing the 1$\rightarrow$3 
dipoles transition which depends on the two four-point conformal ratios ${X^2{Y'}^2\over {X'}^2Y^2}$ 
and ${(x-y)^2(z-z')^2\over X^2 {Y'}^2}$.

In conclusion, I would like to discuss the conformal invariance of the evolution equation (\ref{nlobksym}).
Formally, the light-like Wilson line is invariant under SL(2,C) group of conformal transformations of the transverse plane. 
It looks like one should expect that the corresponding evolution kernel is conformal. 
It should be emphasized, however, that the matrix elements of the light-like Wilson-line 
operators diverge in the longitudinal direction, and when we impose cutoff in rapidity 
we destroy the conformal invariance. The conformal invariance of the ${\cal N}=4$ amplitude 
should be restored after multiplication of the evolution kernel by the coefficient functions 
of the high-energy operator expansion - the so-called ``impact factors''. The study is in progess.


\begin{theacknowledgments}
  This work was supported by DOE grant DE-FG02-97ER41028 and by the
Jefferson Lab Graduate Fellowship. The author thanks the organizers 
for  the warm hospitality received during this workshop.
\end{theacknowledgments}

\bibliographystyle{aipproc}   

\bibliography{sample}

\IfFileExists{\jobname.bbl}{}
 {\typeout{}
  \typeout{******************************************}
  \typeout{** Please run "bibtex \jobname" to optain}
  \typeout{** the bibliography and then re-run LaTeX}
  \typeout{** twice to fix the references!}
  \typeout{******************************************}
  \typeout{}
 }


\end{document}